\documentclass[copyright,creativecommons]{eptcs}
\providecommand{\event}{SR2014}

\usepackage{amssymb}
\usepackage{amsmath}
\usepackage{txfonts}
\usepackage{amssymb}
\usepackage{enumerate}
\usepackage{amsfonts}
\usepackage{times}
\usepackage{mathrsfs}
\usepackage{amscd}
\usepackage{graphicx}
\usepackage{makeidx}
\usepackage{hyperref}

\usepackage{version}
 
\includeversion{full}

\newcommand{\commentout}[1]{}

\newcommand{\CTLsK}{\mbox{CTL$^*$K}}
\newcommand{\ESL}{\mbox{ESL}}

\newcommand{\trans}{\rightarrow}

\newcommand{\R}{{\cal R}}

\newcommand{\until}{U}

\newcommand{\Prop}{Prop}
\newcommand{\SVar}{V\!ar}
\newcommand{\nat}{\mathbb{N}}

      \newtheorem{theorem}{Theorem}

\newcommand{\strat}{\sigma}
\newcommand{\Ags}{\mathit{Ags}} 
\newcommand{\I}{\mathcal{I}} 
\newcommand{\atlop}[1]{\langle\hspace{-2pt}\langle #1 \rangle \hspace{-2pt}\rangle } 
\newcommand{\existsg}[1]{\exists #1.} 
\newcommand{\forallg}[1]{\forall #1.} 
\newcommand{\lid}[2]{\mathtt{e}_{#1}(#2)}

\newcommand{\rimp}{\Rightarrow}

\newcommand{\exploited}{\mathtt{exploited}}  
\newcommand{\cc}{\mathtt{cc}}  
\newcommand{\CCN}{\mathtt{CCN}}  
\newcommand{\done}{\mathtt{done}}

\newcommand{\Env}{E}

\newcommand{\Acts}{\mathit{Acts}}

\newcommand{\be}{\begin{enumerate}} 
\newcommand{\ee}{\end{enumerate}}

\newcommand{\nxt}{\hspace{-1pt}\bigcirc\hspace{-1pt}} 
\newcommand{\always}{\Box} 
\newcommand{\sometimes}{\Diamond}

\newcommand{\G}{{\cal G}} 

\newcommand{\detstrat}{\mathit{det}}

\newcommand{\sgy}{\alpha} 
\newcommand{\unif}{\mathit{unif}} 
\newcommand{\Strat}{\Sigma} 
  
  \newcommand{\Strats}{\Sigma}
    \newcommand{\Cont}{\Gamma}

\newcommand{\ETL}{{\rm ETLK}}

\title{An Epistemic Strategy Logic (Extended Abstract)}

\author{Xiaowei Huang
\institute{The University of New South Wales}
 \and 
Ron van der Meyden
\institute{The University of New South Wales}
}
\def\titlerunning{An Epistemic Strategy Logic}
\def\authorrunning{X. Huang \& R. van der Meyden}
\begin{document}

\maketitle

\begin{abstract}
The paper presents an extension of temporal epistemic logic with operators that quantify over strategies. 
The language also provides a natural way to represent what agents would know were they to be
aware of the strategies  being used by other agents. 
Some examples are presented to motivate the framework, and relationships to 
several variants of alternating temporal epistemic logic are 
discussed. The computational complexity of model checking the logic is also characterized.\end{abstract}

\section{Introduction}

There are many subtle issues  concerning agent knowledge in settings 
where multiple agents act strategically. In the process of understanding these issues, there has been a proliferation 
of modal logics  dealing with epistemic reasoning in strategic settings, e.g., \cite{Schobbens2004,vOJ2005,JA2009}.  
The trend has been for these logics to contain large numbers of operators, each of which  combines several different concerns, such as the 
existence of strategies, and knowledge that groups of agents may have about these strategies. 
We argued in a previous work  \cite{HvdM2014}  that epistemic temporal  logic already has the expressiveness 
required for many applications of epistemic strategy logics, provided that one works 
in a semantic framework in which strategies are explicitly rather than  (as in most alternating temporal epistemic logics) implicitly  represented, and 
makes the minor innovation of including new agents whose local states correspond to the strategies being used by other agents. This gives a more compositional basis for epistemic strategic logic. In the case of imperfect recall strategies and knowledge operators, and a CTL$^*$ temporal basis, 
this leads  to a temporal epistemic strategy logic with a PSPACE complete model checking problem. 

However, some of our 
results
in  \cite{HvdM2014} required a restriction to cases not involving a common knowledge operator, 
because certain notions could not be expressed.
In the present paper, we  develop a remedy for this weakness. 
We propose an epistemic strategy logic which, like~\cite{CHP10,MogaveroMV10}, supports explicit naming and quantification over strategies.
However we achieve this in a slightly more general way: we first generalize temporal epistemic logic to include operators 
for quantification over global states and reference to their components, and then apply this generalization to 
a system that includes strategies encoded in the global states and references these using the ``strategic" agents of  \cite{HvdM2014} . 
The resulting framework can express many of the subtly different notions that have been the
subject of proposals for alternating temporal epistemic logics. In particular, it
generalizes the expressiveness of the  logic in \cite{HvdM2014} but is able to also deal with the common knowledge 
issues that restricted the scope of that work.  
The new logic retains the pleasant compositional capabilities of the prior proposal. 
There is, however, a computational cost to the generalization: the complexity of model checking for the extended language
based on CTL$^*$ is EXPSPACE-complete, a jump over the previous PSPACE-completeness result. However, 
for the fragment based on CTL temporal operators, model checking remains PSPACE-complete.

\subsection{An extended temporal epistemic logic}

We extend temporal epistemic logic with a set of variables $\SVar$, an operator $\existsg{x}$ and a construct $\lid{i}{x}$, 
where $x$ is a variable and $\existsg{x}\phi$ says, intuitively, that there exists in the system a global state $x$ such that $\phi$ holds
at the current point, 
and $\lid{i}{x}$ says that  agent $i$ has the same local state at the current point and at the global state $x$. 
Let $\Prop$ be a set of atomic propositions and let $\Ags$ be a set of agents. 
Formally, the language $\ETL(\Ags,  \Prop,\SVar)$
has syntax given by the grammar: 
$$\phi \equiv p ~|~\neg \phi~|~\phi_1 \lor \phi_2~|
~A\phi ~|~\nxt\phi ~|~\phi_1 \until \phi_2 ~
|~ \existsg{x}\phi~
|~ \lid{i}{x}~
| ~D_G\phi~|~C_G\phi$$ 
where $p \in \Prop$, $x\in \SVar$, $i\in \Ags$, and $G\subseteq \Ags$. 
The construct $D_G\phi$ expresses that agents in $G$ have  distributed knowledge of $\phi$, i.e., could deduce $\phi$ if they pooled their information, 
and $C_G\phi$ says that $\phi$ is common knowledge to group $G$. 
The temporal formulas $\nxt\phi$, $\phi_1 \until \phi_2$, $A\phi$ have the standard meanings from $CTL^*$, i.e., 
$\nxt \phi$ says that $\phi$ holds at the next moment of time, $\phi_1 \until \phi_2$ says that $\phi_1$ holds until $\phi_2$ does,  and
$A\phi$ says that $\phi$ holds in all possible evolutions from the present situation. 
Other operators can be obtained in the usual way, e.g., $\phi_1\land \phi_2 =  \neg (\neg \phi_1\lor \neg \phi_2)$, $\sometimes\phi = (true\until \phi) $, 
 $\always\phi = \neg \sometimes \neg \phi$, etc. 
The universal form $\forallg{x}\phi = \neg \existsg{x} \neg \phi$ expresses that $\phi$ holds
for all global states  $x$. 
For an agent $i \in \Ags$, we  write $K_i \phi$ for $D_{\{i\}} \phi$; 
this expresses that  agent $i$ knows the fact $\phi$. 
The notion of everyone in group $G$ knowing $\phi$ can then 
be expressed as $E_G\phi = \bigwedge_{i\in G} K_i \phi$. 
We write  $\lid{G}{x}$ for 
$\bigwedge_{i\in G} \lid{i}{x}$. 
This says that at the current point, 
the agents in $G$ 
have the same local state as they do at 
the global state named by variable $x$.

The semantics of $\ETL(\Ags,  \Prop,\SVar)$ builds straightforwardly on 
the following definitions used 
in the standard semantics for temporal epistemic logic \cite{FHMVbook}. 
Consider a system for a set of agents $\Ags$. 
A {\em global state} is an element of the set $\G = L_e \times  \Pi_{i\in \Ags} L_i$, 
where 
$L_e$ is a set of states of the environment and each $L_i$ is a  set of {\em local states} for agent $i$. 
A {\em run} is a mapping $r: \nat\rightarrow \G$ giving a global state at each moment of time. 
A {\em point} is a pair $(r,m)$ consisting of a run $r$ and a time $m$. 
An {\em interpreted system} is a pair $\I = (\R, \pi)$, where 
$\R$ is a set of runs and $\pi$ is an  {\em interpretation},  mapping each point $(r,m)$ with $r\in \R$ 
to a subset of  $\Prop$. For $n \leq m$,   write $r[n\ldots m]$ for the sequence $r(n) r(n+1)  \ldots r(m)$. 
Elements of $\R\times \nat$ are called the {\em points} of $\I$. 
For each agent $i \in \Ags\cup \{e\}$, we write $r_i(m)$ for the component of $r(m)$ in $L_i$, and then
define an equivalence relation on points by $(r,m) \sim_i (r',m')$ if 
$r_i(m) = r'_i(m')$. 
We also define $\sim^D_G\equiv \cap_{i\in G}\sim_i$, and $\sim^E_G \equiv \cup_{i\in G} \sim_i$, and $\sim^C_G \equiv (\cup_{i\in G} \sim_i)^*$  for $G\subseteq \Ags$. 
We take $\sim^D_\emptyset$ to be the universal relation on points, and $\sim^E_\emptyset$ and $\sim^C_\emptyset$ to be the 
identity relation. 

To extend this semantic basis for temporal epistemic logic to a semantics for $\ETL(\Ags,  \Prop,\SVar)$, we just need
need to add a construct that interprets variables as global states. 
A {\em context} for an interpreted system $\I$  is a mapping $\Cont$ from $\SVar$ to global states occurring in $\I$. 
We write $\Cont[g/x]$ for the 
context $\Cont'$ with $\Cont'(x) =g$ and $\Cont'(y) = \Cont(y)$ for all variables $y\neq x$. 
The semantics of the language \ETL\  
is given by a relation $\Cont, \I, (r,m) \models \phi$, representing that 
formula $\phi$ holds at point $(r,m)$ of the interpreted system $\I$, relative to context $\Cont$. 
This is defined inductively on the structure of the formula $\phi$, as follows: 
\begin{itemize}
\item $\Cont, \I, (r,m)\models p$ if   $p\in \pi(r,m)$; 
\item
$\Cont, \I, (r,m)\models \neg \phi$ if not $\Cont, \I, (r,m)\models \phi$; 

\item
$\Cont, \I, (r,m)\models \phi\wedge \psi$ if $\Cont, \I, (r,m)\models \phi$ and $\Cont, \I, (r,m)\models \psi$; 

\item
$\Cont, \I, (r,m)\models A\phi$ if  $\Cont, \I, (r',m)\models \phi$ for all $r' \in \R$ with 
$r[0\ldots m] = r'[0\ldots m]$; 

\item
$\Cont, \I, (r,m)\models \nxt\phi$ if  $\Cont, \I,  (r,m+1)\models \phi$; 

\item
$\Cont, \I, (r,m)\models \phi \until \psi$ if  there exists
$m'\!\geq\! m $ such that $\Cont, \I, (r,m')\models \psi$ and  $\Cont, \I, (r,k)\models \phi$ for all $k$ with $m\leq k < m'$;

\item
$\Cont, \I, (r,m)\models \existsg{x}\phi$ if $\Cont[r'(m')/x], \I, (r,m)\models\phi$ for some point $(r',m')$ of $\I$; 

\item
$\Cont,\I, (r,m)\models \lid{i}{x}$ if $r_i(m) = \Cont(x)_i$; 

\item
$\Cont,\I, (r,m)\models D_G\phi$ if 
 $\Cont,\I, (r',m')\models \phi$ for all $(r',m')$ such that 
 $(r',m')\sim^D_G (r,m)$;

\item
$\Cont,\I, (r,m)\models C_G\phi$ if 
$\Cont,\I,(r',m')\models\phi$ for all $(r',m')$ such that $(r',m')\sim_G^C (r,m)$.
\end{itemize}
The definition is standard, except for the constructs  $\existsg{x}\phi$ and $\lid{i}{x}$. 
The clause for the former says that $\existsg{x}\phi$ holds at a point $(r,m)$ if 
there exists a global state $g=r'(m')$ such  
that  $\phi$ holds at the  point $(r,m)$, provided, we interpret $x$ as 
referring to $g$. Note that it is required that $g$ is attained at some point $(r',m')$, 
so actually occurs in the system $\I$. 
The clause for $\lid{i}{x}$ says that this holds at a point 
$(r,m)$ if the local state of agent $i$, i.e., $r_i(m)$, is the same as the 
local state $\Cont(x)_i$ of agent $i$ at the global state $\Cont(x)$ 
that interprets the variable $x$ according to $\Cont$. 

We remark that these novel constructs introduce some redundancy, in that
the set of epistemic operators $D_G$ could be reduced to the ``universal'' operator $D_\emptyset$, 
since  $ D_G\phi \equiv \existsg{x} (\lid{G}{x} \land D_\emptyset (\lid{G}{x} \rimp \phi))$. 
Evidently, given the complexity of this formulation,  $D_G$ remains a useful notation.

\subsection{Strategic Environments}

In order to deal with agents that operate in an environment by strategically choosing their
actions, we introduce a richer type of transition system that models the available actions
and their effects on the state.  An {\em environment} for agents $\Ags$  is a tuple 
$\Env =  \langle S, I, \Acts, \trans, \{O_i\}_{i\in \Ags}, \pi\rangle$, where 
$S$ is a set of states, 
$I$ is a subset of $S$, representing the initial states, 
$\Acts = \Pi_{i\in Ags} \Acts_i$ is a set of 
joint actions, where each $\Acts_i$ is a  nonempty set of actions
that may be performed by agent $i$, 
component 
$\trans \subseteq S \times \Acts \times S$ is a transition relation, 
$O_i: S\rightarrow L_i$ is 
an observation function, 
and 
$\pi: S\rightarrow {\cal P}(\Prop)$ is a propositional assignment. 
An environment is said to be finite if all its components, i.e., $S, \Ags, \Acts_i, L_i$ and $\Prop$ are finite. 
Intuitively, a joint action $a\in \Acts$ represents a choice of action $a_i\in \Acts_i$ for each agent $i\in \Ags$, 
performed simultaneously, and the transition relation resolves this into an effect on the 
state. We assume that $\trans$ is serial in the sense that for all $s\in S$ and 
$a \in \Acts$ there exists $t\in S$ such that $(s,a, t)\in\trans$.

A {\em strategy} for agent $i\in \Ags$ in such an environment $\Env$
is a function $\sgy: S \rightarrow {\cal P}(\Acts_i) \setminus \{ \emptyset\}$, 
selecting a set of actions of the agent at each state.%
\footnote{More generally, 
a strategy could be a function of the history, but we focus here
on strategies that depend only on the final state. 
} 
We call these the actions
{\em enabled} at the state. 
A {\em group strategy}, or {\em strategy profile}, for a group $G$ is a 
tuple $\sgy_G = \langle \sgy_i\rangle_{i \in G}$ where each $\sgy_i$ is a strategy for agent $i$. 
A strategy $\sgy_i$ is {\em deterministic} if $\sgy_i(s)$ is a singleton
for all $s$. 
 A strategy $\sgy_i$ for agent $i$ is {\em uniform} if for all states $s,t$, if $O_i(s) = O_i(t)$, then 
 $\sgy_i(s)  = \sgy_i(t)$. A strategy $\sgy_G = \langle \sgy_i\rangle_{i \in G}$ for a group $G$ 
 is {\em locally uniform (deterministic)} if $\sgy_i$ is uniform (respectively, deterministic) for each agent $i\in G$.
 Given an environment $\Env$,   we write
 $\Strat^\detstrat(\Env)$ for the set of deterministic strategies, 
 $\Strat^\unif(\Env)$ for the set of all 
 locally uniform joint strategies, and  $\Strat^{\unif, \detstrat}(\Env)$ for the set of all deterministic 
 locally uniform joint strategies.    
 
We now define an interpreted system that contains all the possible runs generated 
when agents $\Ags$ behave by choosing a strategy 
 from some set $\Sigma$ of joint strategies in the context of an environment $\Env$. 
One innovation, introduced in \cite{HvdM2014}, 
is that the construction 
introduces new agents 
$\strat(i)$, for each $i\in \Ags$. The observation of $\strat(i)$ is the strategy currently being used by agent $i$. 
Agent $\strat(i)$ is not associated with any actions, and  is primarily for use in epistemic operators, 
to allow reference to what can be deduced 
 were agents to  reason using information about each other's strategies.
For $G\subseteq \Ags$, we write $\strat(G)$ for the set  $\{\strat(i)~|~ i\in  G\}$. 
Additionally, we include an agent $e$ for representing the state of the environment. 
(This agent, also, is not associated with any actions.) 
 
 Formally, given an environment 
 $\Env =  \langle S, I, \Acts, \trans, \{O_i\}_{i\in \Ags}, \pi\rangle$ for agents $\Ags$, 
 where $O_i : S\rightarrow L_i$ for each $i\in \Ags$, 
 and a set $\Strat \subseteq \Pi_{i\in \Ags} \Strat_i$ of joint strategies for the group $\Ags$, 
 we define the {\em strategy space} interpreted system  $\I(\Env,\Strats) = (\R, \pi')$.
The system $\I(\Env,\Strats)$ has global states $\G = S\times \Pi_{i\in \Ags} L_i \times \Pi_{i\in \Ags} \Sigma_i$. 
Intuitively, each global state consists of a state of the environment $E$, a local state for each agent $i$ in $E$, and a strategy for each agent $i$.   
We index the components of this cartesian product by $e$, the elements of $\Ags$ and the elements of $\strat(\Ags)$, respectively. 
We take the set of runs $\R$ of $\I(\Env,\Strats)$ to be the set of all runs $r: \nat \rightarrow \G$
satisfying the following constraints, for all $m\in \nat$ and $i\in \Ags$
\be 
\item $r_e(0) \in I$ and $\langle r_{\strat(i)}(0) \rangle_{i\in \Ags} \in \Strats$, 
\item  $r_i(m) = O_i(r_e(m))$,
\item  $(r_e(m), a,  r_e(m+1)) \in \trans $ for some joint action $a\in \Acts$ such that for all $j\in \Ags$ we have $a_j \in \alpha_j(r_j(m))$, 
where $\alpha_j = r_{\sigma(j)}(m)$, and 
\item $r_{\sigma(i)}(m+1) = r_{\sigma(i)}(m)$. 
\ee
The first constraint, intuitively, says that runs start at an initial state of $E$, and 
the initial strategy profile at time $0$ is one of the profiles in $\Strats$. 
The second constraint states that the agent $i$'s local state  at time $m$ is the observation 
that agent $i$ makes of the state of the environment at time $m$. The third constraint 
says that evolution of the state of the environment is determined at each moment of time by agents choosing 
an action by applying their strategy at that time to their local state at that time. The joint action resulting from these individual choices is then  
resolved into a transition on the state of the environment using the transition relation from $\Env$.  
The final constraint says that agents' strategies are fixed during the course of a run. 
Intuitively, each agent picks a strategy, and then sticks to it. The interpretation $\pi'$ of $\I(\Env,\Strats)$
is determined from the interpretation  
$\pi$
of $\Env$ by taking $\pi'(r,m) = \pi(r_e(m))$ for all points $(r,m)$.

Our epistemic strategy logic is now just  an instantiation of the extended temporal epistemic 
logic in the strategy space generated by an environment. That is, we start with an environment $\Env$
and an associated set of strategies $\Strats$, 
and then work with the language $\ETL(\Ags\cup \strat(\Ags) \cup \{e\}, \Prop, \SVar)$
in the interpreted system 
$\I(\Env,\Strats)$. 
We call this instance of the language $\ESL(\Ags, \Prop, \SVar)$, or just $\ESL$ when the 
parameters are implicit. 

\section{Applications}

In~\cite{HvdM2014}, we proposed a logic 
$\CTLsK(\Ags\cup\strat(\Ags),\Prop)$
extending temporal epistemic logic with strategy agents to allow the reasoning about knowledge and strategy by standard epistemic operators. The language introduced above  is a generalization of the definitions in~\cite{HvdM2014}, to which  we have 
added the constructs $\existsg{x}\phi$ and $\lid{i}{x}$. For formulas without these constructs, the semantics 
of $\ESL$ 
ignores the context $\Cont$, so this 
component of the triple 
$\Cont, \I(\Env, \Sigma), (r,m)$ 
can be removed from the definition, and it collapses to the  
definitions for $\CTLsK(\Ags\cup\strat(\Ags),\Prop)$ in~\cite{HvdM2014}. 

In the system  
$\I(\Env, \Strat)$  
we may refer, 
using distributed knowledge operators $D_G$ where $G$ contains the new strategic agents $\strat(i)$,  to 
what agents would know, should they take into account not
just their own observations, but also information about other agent's strategies. 
For example, the distributed knowledge 
operator $D_{\{i,\strat(i), \strat(j)\}}$ captures what agent $i$ would know, 
taking into account its own strategy and the strategy being used by agent $j$. 
Various applications of the usefulness of these distributed knowledge operators containing strategic agents are given in~\cite{HvdM2014}.
For example, we describe an application 
to {\em erasure policies} 
in computer security in which we write formulas such as 
\[ 
\neg D_\emptyset \neg  (  \done \land  \neg \exploited \land EF \bigvee_{x\in \CCN}  D_{\{ A, \strat(A), \strat(M)\}} (  \cc \neq x)) )
\] 
to state that it is possible for an  attacker $A$ on an e-commerce payment gateway 
to obtain information about a credit card number $\cc$ 
even after the transaction is done, provided that the attacker reasons using knowledge about their own
observations, their own strategy, but also knowledge of the strategy being used by the merchant $M$. 
Here  $\done \land  \neg \exploited$ captures a situation where the transaction is done but the
attacker has not run any exploit, and $D_{\{ A, \strat(A), \strat(M)\}} (  \cc \neq x)$ says that the attacker is able to 
rule out the specific credit card number $x$ from the range of possible values $\CCN$ for the actual 
credit card number $\cc$ (so the attacker has at least one bit of information about the actual credit card number). 
The modality $\neg D_\emptyset \neg$ is used to state that there is a point of the system where the formula holds. 
In particular, since the system builds in all possible strategies for the players, this modality captures a quantification 
over strategies.

In further applications given in~\cite{HvdM2014}, we showed that $\CTLsK(\Ags\cup\strat(\Ags),\Prop)$
can be used to 
express game theoretic equilibria, to 
reason about knowledge-based programs \cite{FHMVbook}, and  that 
many variants of alternating temporal epistemic logics that have been proposed in the 
literature  can be expressed using $\CTLsK(\Ags\cup\strat(\Ags),\Prop)$. We refer the reader to \cite{HvdM2014} for details.

However, we had to make a restriction for some of these expressiveness results 
to formulas that do not contain uses of a common knowledge operator. We now show how the 
extended language of the present paper can remove this restriction. 

Jamroga and van der Hoek \cite{JvdH2004} 
formulate a construct $\atlop{H}^\bullet_{{\cal K}(G)}\phi$ that says, effectively, that there is a strategy  for a group $H$ that another group $G$
knows (for notion of group knowledge ${\cal K}$, which could be $E$ for everyone knows, $D$ for  distributed knowledge, or $C$ for common knowledge) 
to achieve goal $\phi$. The semantics of this construct is given with respect to an environment $\Env$ and a state $s$, and (in outline) is given by 
$E, s\models \atlop{H}^\bullet_{{\cal K}(G)}\phi$ if there exists a uniform strategy $\alpha$ for group $H$
such that for all states $t$ with $s \sim_G^{\cal K} t$, we have that all paths $\rho$ from $t$ that are consistent with $\alpha$ satisfy 
$\phi$. Here $\sim^{\cal K}_G$ is the appropriate epistemic indistinguishability relation on states of $E$. 
We show in \cite{HvdM2014} how $\atlop{H}^\bullet_{{\cal K}(G)}\phi$ can be expressed in $\CTLsK(\Ags\cup\strat(\Ags),\Prop)$
for the cases where ${\cal K}$ is either $E$ or $D$. 

In the case of the operator
$\atlop{H}^\bullet_{C(G)}\phi$, the definition involves the 
common knowledge that a group $G$ of agents would have if they were to reason taking into consideration the strategy being used by another group $H$. 
This does not appear to be expressible using $\CTLsK(\Ags\cup\strat(\Ags),\Prop)$. 
In particular, the formula $C_{G \cup \strat(H)} \phi$ does not give the intended meaning. 
Instead, what needs to be expressed is the greatest fixpoint $X$ of the equation $ X \equiv \bigwedge_{i\in G} D_{\{i\}\cup \sigma(H)} (X \land \phi)$. 
The language $\CTLsK(\Ags\cup\strat(\Ags),\Prop)$ does not include fixpoint operators and it does not seem that  the intended meaning is expressible. 
On the other hand, it can be expressed with $\ESL(\Ags,\SVar,\Prop)$ in a natural way by a formula 
$C_{G}(\lid{\strat(H)}{x} \rimp \phi)$, 
which says that it is common knowledge to the group $G$ that $\phi$ holds if the 
group $H$ is running the strategy profile capture by the variable $x$. 
Using this idea, the 
 construct $\atlop{H}^\bullet_{C(G)}\phi$ can be represented with $\ESL$  as 
$$ \existsg{x}  C_{G}(\lid{\strat(H)}{x} \rimp \phi)~.$$
(We remark that a carefully stated equivalence result requires an appropriate treatment of 
initial states in the environment $\Env$. We refer to~\cite{HvdM2014} for details.) 
Applying similar ideas, $\ESL$ can also be used to 
eliminate, from the results on reasoning about knowledge-based programs presented in  \cite{HvdM2014},
the restriction to knowledge-based programs not containing common knowledge operators.

\section{Model Checking} 

\newcommand{\stratdesc}{sd}
\newcommand{\stratconf}{\gamma}

Since interpreted systems are always infinite objects, we use environments to give a finite input for the model checking problem. 
For an environment $\Env$, a set of strategies $\Strat$ for $\Env$, and a context $\Cont$ for  
$ \I(\Env, \Strat)$, 
we write $\Cont, E,\Strat\models \phi$ if 
$\Cont, \I(\Env, \Strat),(r,0)\models \phi$ 
for all runs $r$ of $\I$ of 
$\I(\Env, \Strat)$. 
(Often, the formula $\phi$ will be a sentence, i.e., will have all variables $x$ in the scope of an operator $\exists x$. 
In this case the statement $\Cont, \Env,\Strat\models \phi$ is independent of $\Cont$ and we may write simply $\Env,\Strat\models \phi$) 
The model checking problem is to  determine whether $\Cont, \Env,\Strat\models \phi$ for a finite state environment $\Env$, a set $\Strats$ of strategies 
and a context $\Cont$, where  $\phi$ is an $\ESL(\Ags,\SVar,\Prop)$ formula. 

For generality, we abstract $\Strats$ to a paramaterized 
class such that for each environment $\Env$, the set $\Strats(\Env)$ is a set of strategies for $\Env$. 
We say that the parameterized class $\Strats(\Env)$ is {\em PTIME-presented}, 
if it is presented by means of an  algorithm that runs in time polynomial in the size of $E$ and 
verifies if a given strategy $\alpha$ is in $\Strats(E)$. For example, the class $\Strats(E)$ of all strategies for 
$E$ can be PTIME-presented, as can $\Strats^\unif(E)$,  $\Strats^{\detstrat}(E)$ and  $\Strats^{\unif,\detstrat}(E)$. 

A naive model checking  algorithm would construct a transition system 
over the set of states $S\times \Strats(\Env)$, where $S$ is the set of states of $\Env$,  
then apply model checking techniques 
on it. 
Note that  a joint strategy for an environment $\Env$ can be represented in space $|S|\times |\Acts|$.
Thus, the size of $S\times \Strats(\Env)$ is exponential  as a function of the size of $\Env$. 
This means that the naive procedure requires exponential space. 
This indeed turns out to be the complexity of model checking the logic. 
However, it is possible to do better than this 
provided we restrict to the CTL-based fragment of the language. This is 
the fragment in which the temporal operators occur only in the forms
$A\nxt \phi$, $\neg A\neg\nxt \phi$, $A\phi_1 \until \phi_2$, and $\neg A\neg \phi_1 \until \phi_2$.

\begin{theorem}
Let $\Strats(\Env)$  be a PTIME presented class of strategies for environments $\Env$. 
The complexity of deciding, given an environment $\Env$, an \ESL\ formula $\phi$ 
and a context $\Cont$ for the free variables in an $\ESL$  formula $\phi$ relative
to $\Env$ and $\Strats(\Env)$,  whether $\Cont, \Env,\Strat(\Env)\models \phi$, is EXPSPACE-complete. 
For the restriction of the problem to $\phi$ in the CTL-based  fragment, the complexity is 
PSPACE-complete. 
\end{theorem}

\section{Conclusions}

{\em Hybrid} logic \cite{BS98} is an approach to the extension of modal logics that 
uses ``nominals", i.e., propositions $p$ that hold at a 
single world. These can be  used in combination with operators
such as  $\exists p$, which marks an arbitrary world
as the unique world at which nominal $p$ holds. Our construct $\exists x$ is 
closely related to the hybrid construct  $\exists p$, but we work in a setting
that is richer in both syntax and semantics than previous works. 
There have been a few works using hybrid logic ideas in the context of epistemic logic \cite{Hansen11,Roy09a} 
but none are  concerned with temporal logic. Hybrid temporal logic has seen a larger
amount of study \cite{BozzelliL10,FranceschetRS03,FranceschetR06,SchwentickW07}, 
with variances in the semantics used for the model checking problem. 

We note that if we were to view the variable $x$ in our logic as a propositional constant, 
it would be true at a set of points in the system $\I(\Env, \Strats)$, hence not a nominal in that system.
Results in \cite{BozzelliL10}, where a hybrid linear time temporal logic
formula is checked in all paths in a given model, suggest that 
a variant of $\ESL$ in which $x$ is treated  as a nominal in $\I(\Env, \Strats)$
would have a complexity of model checking at least non-elementary, compared to 
our EXPSPACE and PSPACE complexity results. 

Our model checking result seems to be more closely related to the a result in \cite{FranceschetR06} that 
model checking a logic HL$(\exists, @, F,A)$ is PSPACE-complete. 
Here $F$ is essentially a branching  time future operator and $A$ is a universal operator (similar to 
our $D_{\emptyset}$), the construct $@_p\phi$ says that $\phi$ holds at the world marked by the 
nominal $p$, and $\exists p(\phi)$ says that $\phi$ holds after marking some world by $p$. 
The semantics in this case does not unfold the model 
into either a tree or a set of linear structures before checking the formula, 
so the semantics of the hybrid existential $\exists$ is close to our idea
of quantifying over global states. Our language, however, has a richer set of 
operators, even in the temporal dimension, and introduces the strategic dimension in the 
semantics. It would be an interesting question for future work to 
consider fragments of our language to obtain more precise statement of the
relationship with hybrid temporal logics. 

Strategy Logic \cite{CHP10} is a (non-epistemic) generalization of  ATL for perfect information 
strategies in which strategies may be explicitly named and quantified. 
Work on identification of more efficient variants 
of quantified strategy logic includes \cite{MogaveroMV10}, which formulates a variant 
with a 2-EXPTIME-complete model checking problem. 
In both cases, strategies are perfect recall strategies, rather than the imperfect recall 
strategies that form the basis for our PSPACE-completeness result for model checking. 
The exploration of our logic over such a richer space of strategies is an interesting
topic for future research.

\bibliographystyle{eptcs}
\bibliography{atl}

\end{document}